\documentclass[12pt]{article}
\usepackage{hyperref}
\usepackage{cite}
\usepackage{color}
\usepackage{graphicx}
\usepackage{amsmath}
\usepackage{amssymb}
\usepackage{xspace}
\usepackage{oubraces}
\usepackage[normalem]{ulem}
\usepackage[titletoc]{appendix}
\usepackage{tikz}
\usetikzlibrary{decorations.pathmorphing}

\makeatletter
\@addtoreset{equation}{section}

\makeatletter
\renewcommand\section{\@startsection {section}{1}{\z@}%
                                   {-3.5ex \@plus -1ex \@minus -.2ex}
                                   {2.3ex \@plus.2ex}%
                                   {\normalfont\large\bfseries}}
\renewcommand\subsection{\@startsection{subsection}{2}{\z@}%
                                     {-3.25ex\@plus -1ex \@minus -.2ex}%
                                     {1.5ex \@plus .2ex}%
                                     {\normalfont\bfseries}}

\def\baselinestretch{1.2}
\newcommand{\be}{\begin{equation}}
\newcommand{\ee}{\end{equation}}
\newcommand{\beq}{\begin{eqnarray}}
\newcommand{\eeq}{\end{eqnarray}}

\newcommand{\gone}[1]{{}}

\newcommand{\bmat}{\left(\begin{array}}
\newcommand{\emat}{\end{array}\right)}

\def\yzero{\smash{\hbox{$y\kern-4pt\raise1pt\hbox{${}^\circ$}$}}}

\def\-{\hphantom{-}}

\def\s2{\frac{1}{\sqrt2}}

\def\IF{\relax{\rm I\kern-.18em F}}
\def\II{\relax{\rm I\kern-.18em I}}

\def\Dsl{\,\raise.15ex\hbox{/}\mkern-13.5mu D}

\newcommand{\eq}[1]{(\ref{#1})}
\newcommand{\ket}[1]{\vert #1 \rangle}

\catcode`\@=11   
\newdimen\@rotdimen
\newbox\@rotbox  

\def\@vspec#1{\special{ps:#1}}
\def\@rotstart#1{\@vspec{gsave currentpoint currentpoint translate
   #1 neg exch neg exch translate}}
\def\@rotfinish{\@vspec{currentpoint grestore moveto}}
\def\@rotr#1{\@rotdimen=\ht#1\advance\@rotdimen by\dp#1%
   \hbox to\@rotdimen{\hskip\ht#1\vbox to\wd#1{\@rotstart{90 rotate}%
   \box#1\vss}\hss}\@rotfinish}
\def\@rotl#1{\@rotdimen=\ht#1\advance\@rotdimen by\dp#1%
   \hbox to\@rotdimen{\vbox to\wd#1{\vskip\wd#1\@rotstart{270 rotate}%
   \box#1\vss}\hss}\@rotfinish}%
\def\@rotu#1{\@rotdimen=\ht#1\advance\@rotdimen by\dp#1%
   \hbox to\wd#1{\hskip\wd#1\vbox to\@rotdimen{\vskip\@rotdimen
   \@rotstart{-1 dup scale}\box#1\vss}\hss}\@rotfinish}%
\def\@rotf#1{\hbox to\wd#1{\hskip\wd#1\@rotstart{-1 1 scale}%
   \box#1\hss}\@rotfinish}%
\def\rotate{\@ifnextchar[{\@rotate}{\@rotate[l]}}
\def\@rotate[#1]#2{\setbox\@rotbox=\hbox{#2}\@nameuse{@rot#1}\@rotbox}

\catcode`\@=12

\begin{document}
\begin{titlepage}
\begin{flushright}
\end{flushright}

\vfil

\begin{center}

{\bf \Large
Complexity Is Simple!
}

\vfil

William Cottrell$^{1}$,  Miguel Montero$^{2}$\\

\vfil

{}$^1$ Institute for Theoretical Physics Amsterdam, University of Amsterdam\\ 1098 XH Amsterdam, The Netherlands

{}$^2$Institute for Theoretical Physics and
Center for Extreme Matter and Emergent Phenomena,\\
Utrecht University, Princetonplein 5, 3584 CC Utrecht, The Netherlands

\vfil

\end{center}

\begin{abstract} 
  \noindent In this note we investigate the role of Lloyd's computational bound in holographic complexity.  Our goal is to translate the assumptions behind Lloyd's proof into the bulk language. In particular, we discuss the distinction between orthogonalizing and `simple' gates and argue that these notions are useful for diagnosing holographic complexity.  We show that large black holes constructed from series circuits necessarily employ simple gates, and thus do not satisfy Lloyd's assumptions.  We also estimate the degree of parallel processing required in this case for elementary gates to orthogonalize.    Finally, we show that for small black holes at fixed chemical potential, the orthogonalization condition is satisfied near the phase transition, supporting a possible argument for the Weak Gravity Conjecture first advocated in \cite{Brown:2015bva}.           
\end{abstract}
\vspace{0.5in}

\end{titlepage}
\renewcommand{\baselinestretch}{1.05}  

\tableofcontents

\section{Introduction}

The action, $\mathcal{A}$, and volume, $\mathcal{V}$, of $AdS_{d+1}$ wormholes have emerged as interesting new holographic observables.  Given the simple and universal character of these quantities, it seems reasonable to conclude that there must be natural field theory dual.  Indeed, as pointed out in the recent literature  \cite{Brown:2015bva, Brown:2015lvg}, computational complexity, $\mathcal{C}$, seems to be a promising candidate, having the same gross features as the target observables.  This connection has led both to a `complexity equals volume' ($\mathcal{C} = \mathcal{V}$) conjecture \cite{Stanford:2014jda} and more recently to `complexity equals action' ($\mathcal{C} = \mathcal{A}$) \cite{Brown:2015bva, Brown:2015lvg}.  If the relation between complexity and bulk action and/or volume proves true, it would be a remarkable new addition to the holographic dictionary.

One exciting feature of such a connection would be that it more directly ties bulk observables to concepts in quantum information theory (QIT).  In principle then, one could hope to `import' ideas from QIT to make statements that would not be manifest from the bulk gravity description.  Potentially, it may even be possible to constrain the set of allowed bulk theories and thus become a powerful weapon in the crusade to banish the  `Swampland'  \cite{ArkaniHamed:2006dz,delaFuente:2014aca,Rudelius:2014wla,Rudelius:2015xta,Montero:2015ofa,Brown:2015iha,Bachlechner:2015qja,Hebecker:2015rya,Brown:2015lia,Junghans:2015hba,Palti:2015xra,Heidenreich:2015nta,Kooner:2015rza,Heidenreich:2015wga,Ibanez:2015fcv,Montero:2016tif,Heidenreich:2016aqi,Hebecker:2016dsw,Saraswat:2016eaz,Herraez:2016dxn,Ooguri:2016pdq,Cottrell:2016bty,Hebecker:2017wsu,Palti:2017elp,Hebecker:2017uix,Klaewer:2016kiy,Ibanez:2017oqr,Hamada:2017yji,Montero:2017yja,Montero:2017mdq,Ibanez:2017vfl,Lust:2017wrl}.  

An application of this sort was suggested already in \cite{Brown:2015bva}, where it was noted that Lloyd's bound \cite{2000Natur.406.1047L} for computational speeds seemed to imply a natural upper limit on the rate of complexification.  After fixing one adjustable pre-factor in the definition of complexity,  \cite{Brown:2015bva}  found that indeed neutral AdS black holes in various dimensions saturate the Lloyd computation bound if one assumes the `complexity = action' conjecture. The bound is also non-trivially satisfied for rotating as well as small charged black holes, with saturation occurring at extremality. 

However, one notable exception occurs for large near-extremal $AdS$ black holes, which  violate the Lloyd complexification bound under the `$\mathcal{C} = \mathcal{A}$' hypothesis.  This is a tantalizing observation since there is another conjecture, the Weak Gravity Conjecture (WGC), which tells us that these black holes are not supposed to exist as stable states in any UV completion.  According to the standard lore of Weak Gravity, there should exist some light particle in the spectrum which mediates the decay of near-extremal black holes via Schwinger production.  This process would then salvage the putative restriction due to Lloyd's bound.

This note was partially motivated by the converse question, ``Can Lloyd's bound be leveraged into a `proof' of the Weak Gravity Conjecture using $\mathcal{C} = \mathcal{A}$?".  This goal is far from being achieved.  Here, instead, we seek to expose as clearly as possible the `missing steps' in establishing a proof of this sort. Although our attitude is sympathetic to both $\mathcal{C} = \mathcal{A}$ and WGC, we would like to draw particular attention to one glaring hole in the application of Lloyd's bound to holography, namely, the fact that the bound is only applicable to quantum circuits whose gates implement `orthogonalizing' transformations, i.e., gates which map a wavefunction to one orthogonal to itself.  In the absence of this crucial assumption, no useful computation bound is known.  Moreover, when this assumption is violated, trivial examples of systems with arbitrarily high computation rates may be produced.  

The crucial question is then whether or not the $\mathcal{C} = \mathcal{A}$ conjecture is consistent with an orthogonalizing gate structure.  In this note, we will investigate this issue and show that only for small black holes near a Hawking-Page phase transition can the orthogonalizing assumption be achieved for a series circuit model.  For large black holes, we place restrictions on the circuit that are necessary (but not sufficient) to achieve consistency between $\mathcal{C} = \mathcal{A}$ and the orthogonalizing assumption made in Lloyd's bound.  We will show that for a series circuit model in the canonical ensemble, the gates must be `simple', by which we shall mean that they barely rotate the wavefunction at all. In this case, the original derivation of Lloyd's bound is inapplicable.  Moreover, we show that holographic gates for large black holes must be simple unless the degree of parallelization exceeds a certain bound depending on the charge and $AdS$ radius.   We also provide some general arguments that a smooth notion of complexity demands that the typical gate is `simple' and thus not immediately suitable for an application of Lloyd's bound.   It is hoped that these observations serve to outline the hurdles towards making rigorous complexification bounds in holography.

The remainder of this note is organized as follows.  In Section \ref{qip} we will review some some preliminary statements about quantum information theory, including computational complexity, Margolus-Levitin theorem and Lloyd's bound.  In Section \ref{bhc} we will review the `$\mathcal{C} = \mathcal{A}$' conjecture, and then use this to deduce some properties of the gates. Finally, we offer our conclusions in Section \ref{conclusions}.

\section{Quantum Information preliminaries}
\label{qip}

We will start by reviewing various QI ingredients which will be relevant to our discussion: The definition of quantum complexity, the Margolus-Levitin theorem and orthogonalization times. We will also discuss Lloyd's bound \cite{2000Natur.406.1047L} as a complexity bound and the implications of serial versus parallel computation. Finally, we introduce the notion of a \emph{simple} gate, which will be central to our discussion later on. 

\subsection{Defining complexity}
As discussed in the Introduction, there is mounting evidence that some notion of complexity can be very useful in understanding several aspects of black hole physics, such as the late-time dynamics or the presence/absence of firewalls \cite{Susskind:2015toa}. The precise definition of holographic complexity is somewhat elusive, so we will start by reviewing the standard notion of quantum complexity $\mathcal{C}_\epsilon$ for an $n$-qubit system and then describe the slight modification needed for holographic applications.  We will follow \cite{2008arXiv0804.3401W}, which defines complexity of a quantum circuit and also describe the closely related notion of complexity of a state (see e.g. \cite{Jefferson:2017sdb}).

The standard definition of complexity is as follows. Pick a reference state $\ket{0}$, and a {\it{fixed}} finite universal set of elementary gates, so that an arbitrary unitary operator can be approximated to any desired accuracy by a finite product of elementary gates. For an $n$-qubit system, \cite{2008arXiv0804.3401W,Nielsen:2011:QCQ:1972505} provides an example of such a set: It is composed of a Hadamard gate for each pair of qubits, a phase shift gate, and a Toffoli gate\footnote{For applications to holography, one might instead want to build the elementary gate set from more physical quantities such as Hamiltonian density or other single-trace operators.}. With suitable UV and IR cutoffs applied, we can always represent the system of interest in terms of a countable number of qubits.  Simply take the discrete spectrum of energy eigenstates and label them in binary.  With this caveat, we can imagine that this discussion of the $n$-qubit system is general.

Under these assumptions, we may define $\mathcal{F}_k$ as the set of vectors in the Hilbert space that are obtained by acting on $\ket{0}$ with a product of $k$ elementary gates. Now fix a tolerance $\epsilon>0$, and let 
\begin{align}d(\ket{a},\ket{b})=\arccos\left(\frac{\vert\langle a\ket{b}\vert}{\langle a\ket{a}\langle b\ket{b}}\right)\label{fubini}\end{align}
be the Fubini distance between states. The complexity $\mathcal{C}_\epsilon$ of a general state $\ket{\Psi}$ is 
\begin{align}\mathcal{C}_\epsilon(\ket{\Psi})=\text{min}(\left\{k\, :\, \exists \ket{f}\in \mathcal{F}_k\ \text{s.t.}\ d(\ket{f},\ket{\Psi})<\epsilon\right\}).\label{ssa}\end{align}
In other words, $\mathcal{C}_\epsilon$ is the smallest number of gates needed to get within a tolerance of $\epsilon$ from the given state. 

It is apparent that this definition depends on the choice of gates and it is not clear at the moment how to extract universal gate-independent properties.  One immediate property is that complexity decreases under inclusion.  I.e., given two sets of gates $\mathbf{G}$ and $\mathbf{G'}\supset \mathbf{G}$, we have $\mathcal{C}_\epsilon^{\mathbf{G}}\geq \mathcal{C}_\epsilon^{\mathbf{G}'}$. However, the general dependence of this definition on the set $\mathbf{G}$ is not well understood.  

It is also obvious that complexity as defined above is far from a smooth function on the Hilbert space as moving the state away a little bit can mean that a very different $\ket{f}$ becomes the closest one.  To illustrate this point, we will compute the complexity $\mathcal{C}_\epsilon$ in a single qubit system for an arbitrary state of the form
\begin{align}\ket{\theta}= \cos\theta\ket{0}+\sin\theta\ket{1},\label{ss}\end{align}
i.e. a Bloch sphere meridian, for some values of $\epsilon$. To do this, we will compute all the $\mathcal{F}_k$ until some $k_\text{max.}$, and then compute \eq{ssa} with this cutoff. If the set is empty, we know that $\mathcal{C}_\epsilon(\ket{\Psi})\geq k_\text{max.}$. For a single qubit system, we only need two gates to form a universal set, which we take to be the square root of the Hadamard and $\pi/8$ gates \cite{Nielsen:2011:QCQ:1972505}.

\begin{figure}[!htb]\begin{center}
\includegraphics{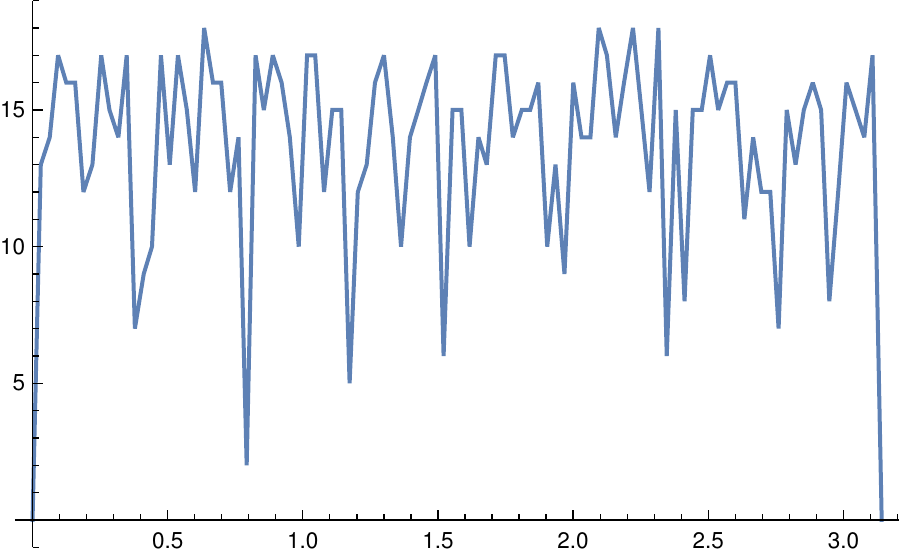}\end{center}
\caption{Complexity as a function of $\theta$ for the state \eq{ss}. The plot was done with $k_{\text{max}}=19$ and  $\epsilon=0.015$.  We see that complexity is a fairly discontinuous thing.}
\label{fig1}
\end{figure}

Figure \ref{fig1} shows an illustrative plot. The take home message is (as was to be expected) that the  standard notion of complexity of a state is a very discontinuous function as one moves around the Hilbert space. (In Appendix \ref{somebound} we will derive a `worst case scenario' bound for this situation.) Furthermore, typical states\footnote{With respect to the measure (\ref{fubini}).} have high complexity. This is also intuitively obvious since the number of states reached in a finite number of steps is merely a finite subset of a continuum.  Moreover the number of states in $\mathcal{F}_k$ is exponential in $k$, meaning that high complexity states are vastly dominate the low complexity ones.

In the traditional definition of complexity there is some room to alter the set of gates used but they are nevertheless fixed at the outset and independent of the tolerance.  This inexorably leads to a very discontinuous notion of complexity.  However, the ultimate interest is in defining a notion of complexity that mirrors smooth universal gravitational observables\footnote{We use this word here in the loose sense.} such as the action or volume.  To achieve this, one must allow the gate set to change with the the tolerance $\epsilon$.  In other words, as the tolerance becomes smaller, we should also choose our gates to make finer changes to the wavefunction.  This is the strategy employed in \cite{Jefferson:2017sdb, Chapman:2017rqy}, where the gates are chosen $G_{i} =e^{i \epsilon Q_{i}}$, for some fixed set $Q_{i}$.  In other words, all gates are chosen to approach the identity as the tolerance vanishes.  

This motivates the introduction of a notion of `simple gate', also discussed \cite{Susskind:2014moa}.  More precisely, we will define a gate to act simply on the state $|\psi\rangle$ if  $\langle \psi | G_{i} |\psi \rangle \sim 1- \mathcal{O}(\epsilon)$, where $\epsilon$ is the tolerance\footnote{
More generally, one could consider a requirement like $\langle \psi | G_{i} |\psi \rangle \sim 1- \mathcal{O}(\epsilon^{\alpha})$ where $\alpha>0$ is an appropriately chosen exponent.  It is clear that $\alpha$ must be positive so as to avoid the highly sporadic behavior illustrated in Figure \ref{fig1}.  On the other hand, a very large value of $\alpha$ would imply that the complexification rate slows to zero in the limit where the cutoff is removed.  The fact that the distance between states $G_{i} |\psi\rangle$ and $|\psi\rangle$ naturally scales with $\epsilon$ makes the choice $\alpha = 1$ a natural one.  However, there may be other possibilities worth considering.}.  Although this definition is slightly outside the original framework described in \cite{2008arXiv0804.3401W}, it is nevertheless in harmony with recent attempts to represent a quantum circuit by taking the continuum limit of the tensor network preparing the state \cite{Caputa:2017yrh}.  From the tensor network perspective, each node in the network contributes a gate like $e^{v \mathcal{O}}$, where $\mathcal{O}$ is a hermitian operator and `$v$' is roughly the infinitesimal volume associated with the gate.  Thus, these gates are `simple' by our criteria, as are any gates that are reasonably associated with holographic complexity.

We shall return to the notion of simple computations in the next section.  The results in the remainder of this Section are independent of this issue and will apply for any choice of $\mathbf{G}$.

\subsection{Margolus-Levitin}
\label{MLS}
We would now like to explore the assumptions underlying the application of Lloyd's bound to complexity. To do this, we first need to discuss the main ingredient in deriving Lloyd's bound, which is the Margolus-Levitin (ML) theorem \cite{Margolus:1997ih}. This theorem sets an upper bound on the time that it takes a quantum system to reach an orthogonal state. The proof is simple enough to reproduce in detail. Let $S(t)=\langle\Psi (0)\ket{\Psi(t)}$ be the overlap as a function of time and also define $\tilde{S}(t) = e^{- i E_{0} t} S(t)$ where $E_{0}$ is the ground state energy. An orthogonal state is reached when $|S| = |\tilde{S}| = 0$, but it is slightly more convenient to work with the latter.  One has
\begin{align}\text{Re}(\tilde{S}(t))=\sum_{n=0}^\infty\vert c_n\vert^2 \cos(\Delta E_{n} t),\end{align}
where the $\Delta E_{n} = E_{n} - E_{0}\ge 0$. Then, using the inequality $\cos x\geq 1-\frac{2}{\pi} (x+\sin x)$, valid for $x\geq 0$, they bound the second term in the above equation as
\begin{align}\text{Re}(\tilde{S}(t))\geq  \sum_{n=0}^\infty\vert c_n\vert^2\left(1-\frac{2\Delta E_n}{\pi} t-\frac{2}{\pi}\sin(\Delta E_n t)\right)=1-\frac{2\Delta E}{\pi}t+\frac{2}{\pi}\text{Im}(\tilde{S}(t)).\label{dwde}\end{align}
where $\Delta E \equiv \langle E_{n} - E_{0}\rangle$. Now, ML look for exact orthogonality, setting $\tilde{S} =S=0$ above and thus obtaining the bound. If we are interested only in approximate orthogonality, we can write $\tilde{S}(t)=\epsilon\, e^{i\alpha}$, and rearrange \eq{dwde} to get
\begin{align}t\geq \frac{\pi}{2\Delta E}\left[1+\epsilon\left(\frac{2}{\pi}\sin\alpha-\cos\alpha\right)\right].\end{align}
Since the minimum of $\frac{2}{\pi}\sin\alpha-\cos\alpha$ is $-\sqrt{1+(2/\pi)^2}$, we find that we need a time
\begin{align}t_\epsilon\geq \frac{\pi}{2\Delta E}\left(1-\epsilon\, \sqrt{1+\frac{4}{\pi^2}}\right)\label{tml0}\end{align}
to reduce the modulus of the overlap below $\epsilon$.  For $\epsilon=0$, we recover the Margolus-Levitin bound
\begin{align}\tau_{\text{orth.}}\geq \tau_{\text{ML}} \label{tml}\end{align}
where we have defined the Margolus-Levitin time as $\tau_{\text{ML}}\equiv \frac{\pi}{2\Delta E}$.  

The bound is also true for large $\epsilon$, but for $\epsilon > \pi/\sqrt{4+\pi^{2}}$ we merely recover the trivial result that $t_{\epsilon} >0$.  In the regime where $|S| = \epsilon \equiv 1-\tilde{\epsilon}$ for $\tilde{\epsilon} \ll 1$, we know of no useful bound, but one may instead derive an equality $t_{\epsilon} \rightarrow \sqrt{2 \tilde{\epsilon}}/\Delta E_{rms}$, where $\Delta E_{rms}\equiv \langle H^{2}\rangle - \langle H\rangle^{2}$.  This is obtained simply by expanding $U = e^{i H t}$ to quadratic order, which is valid for sufficiently small times. A priori there is no hierarchy between $\Delta E_{rms}$ and $\Delta E$, so no obvious route to derive a bound analogous to Lloyd's in this regime.

The discussion for $\epsilon\neq0$ means that the ML result is robust in the sense that it doesn't matter if we talk about approximate or exact orthogonalization.  This is not a trivial point: it is easy to come up with examples in which approximate orthogonalization to any desired precision is achieved long before the exact orthogonalization time $\tau_{\text{orth.}}$. As an example, consider a system of $n$ independent qubits. We act on each of them with Hamiltonian $H=1+\sigma_x$, which gives the state $\ket{0}$ an expected energy of $1$, and takes $\ket{0}$ to $\ket{1}$ in a time $t=\pi/2$, so its exact orthogonalization time is $n$ times the ML time $\tau_{\text{ML}}=\pi/(2n)$, so it satisfies the bound \eq{tml} amply.

The modulus of the overlap with the initial state is simply
\begin{align} \vert S(t)\vert= \vert\cos(t)\vert^n.\end{align}
Now take any $t_1<n \tau_{\text{ML}}$, write $t_1= \alpha \tau_{\text{orth.}}$ with $\alpha<1$.  For large enough $n$, the overlap
\begin{align} \vert \text{overlap}(t_1)\vert= \vert\cos\left(\frac{\pi\alpha}{2}\right)\vert^n\end{align}
is, for any fixed $\alpha$, as small as we want. So we have very approximate orthogonalization far before the actual orthogonalization time of the system. 

On the other hand, it is still true that if we take instead $t_2= \alpha \tau_{\text{ML}}$, a time smaller than the ML time, then the overlap  $\vert \text{overlap}(t_2)\vert$ goes to 1 for any fixed $\alpha$. So even approximate orthogonalization  cannot happen much faster than the ML time,  in accordance with our general result (\ref{tml0}).  The fact that the bound \eq{tml0}  holds also for any $\epsilon>0$ means that in any case it is never possible to achieve approximate orthogonalization long before $\tau_{\text{ML}}$. 

\subsection{Lloyd's bound and simplicity}
\label{lbas}
Armed with the ML theorem, we can now discuss Lloyd's bound and its relationship to complexity, which is one of the main points of this note. 

According to Lloyd \cite{2000Natur.406.1047L}, the ML theorem can be used to derive a fundamental upper limit in computation speed for a {\it{classical}} computer. A computer is supposed to take an initial state $\ket{0}$ to a final one $\ket{\Psi}$, via successive application of logic gates. Let us now focus on a single such step, which results in the addition of a single fundamental gate  $G\in \mathbf{G}$. The gate takes some time $\Delta t$ to perform its task. In other words, $G$ is implemented by a Hamiltonian action, $G=\exp(i H_{\text{gate}} \Delta t)$, with an energy $E=\langle H_{\text{gate}}\rangle$. Higher $E$ means that it takes less time for the gate to perform its task.

Now, Lloyd makes a crucial extra assumption: Because he is interested in limits for \emph{classical} computers, the gates $G$ he considers are all quantum-mechanical implementations of classical logic gates acting on `classical' states. In particular, they all send classical states (which are not qubit superpositions) to classical states.  Since any two different classical states are orthogonal, all the $G$ are chosen to evolve the quantum states they encounter to orthogonal ones, i.e., $\langle G \rangle= 0$ at any step in the computation. But then, the ML theorem \eq{tml} means that each gate must take at least the minimum time to perform its task. For each time step we thus have\footnote{In the following equations we will always take the ground state energy to be zero without loss of generality.}
\begin{align}\Delta t\geq \frac{\pi}{2}\frac{1}{E}
\label{1swq}
\end{align}
where $E$ may be interpreted as the instantaneous energy of the system,  which we assume constant throughout the computation\footnote{There is no obstruction to considering a time-varying energy, though it would add nothing to this discussion.}. We can trivially  rearrange \eq{1swq} to get an upper bound on the instantaneous number of operations per unit time,
\begin{align}\frac{1}{\Delta t}\leq \frac{2 E}{\pi} \label{lloyd0}\end{align}
Equation \eq{lloyd0} is Lloyd's bound, which is an elementary consequence of ML.  \cite{Brown:2015bva,Brown:2015lvg}. 

We may derive Lloyd's bound \eq{lloyd0} under a slightly more general set of assumptions.  Rather than assuming a sequential application of gates we may work in parallel.   For instance, over time step $\Delta t$ we apply the gate:
\be
\ket{0} \rightarrow G_{1}(\Delta t)G_{2}(\Delta t) .... G_{n}(\Delta t)\ket{0}
\ee
where we assume that the $n$ fundamental gates $G_{i}$ are commuting.  In this step, the number of gates in the circuit increases by $n$ while the total energy is $E = \sum E_{i}$.  It is not hard to show that $\Delta t \ge \frac{\pi}{2 \min E_{i}} \ge \frac{ \pi n}{2 E}$.  In other words, the time to add a single circuit, $\Delta t/n$, is still constrained by Lloyd. Combined with our previous argument, this proves Lloyd's bound for  any reasonable ``classical'' computer, which involves a series of steps each of which may contain several elementary commuting gates.

The assumption of a series circuit makes sense for a physical system with interactions that are sparse with respect to the time-scale of observation, while the parallel picture is more applicable to situations where the dynamics cannot be easily disentangled in time. It has been argued that the serial description is the one that applies naturally to black holes \cite{2000Natur.406.1047L} since the optimal bit flip time is comparable to the light crossing time. We will return to this point in Section \ref{bhc}.

We should emphasize that \eq{lloyd0} is derived for \emph{classical} computers. Technically, this is reflected in the fact that the gates $G_i$ are orthogonalizing the states they encounter. For a quantum computer, there is no need to work with orthogonal states; the whole point of quantum computation is to take advantage of quantum superpositions.  As a result, one does not expect a bound like \eq{lloyd} to hold in general, and in fact one can easily find violations when simple gates are used. This was recently illustrated in the works \cite{Jordan:2017vqh,2017arXiv170105550S}, where arbitrarily large complexification rates are achieved in particular examples - in the latter case, in a system with one qubit only.  To illustrate this point, assume we have a reference state $\ket{0}$ and any other state $\ket{\Psi}$ and suppose we have unitary evolution described by
\begin{align}U(t)\ket{0}=\frac12\left((e^{2 i E t}+1)\ket{0}+(e^{2 i E t}-1)\ket{\Psi}\right)\label{celloyd}\end{align}
This maps $\ket{0}$ to $\ket{\Psi}$ in a time $\tau_{\text{ML}}$, no matter how high the complexity of $\ket{\Psi}$ is. The point that one can violate \eq{lloyd0} using non-orthogonalizing transformations is not new; see for instance \cite{2002PhLA..302..291D}.   If we wanted to derive a `Lloyd-like' bound for a generic quantum computer built from simple gates, one would need the kind of bound discussed after (\ref{tml}) with $\tilde{\epsilon} \ll 1$. We offer a candidate in Appendix A, although the resulting constraint is much weaker than Lloyd's bound.

\subsection{Relation to complexity}
\label{rtc}
Equation \eq{lloyd0} is a statement about computers; we now translate it to a statement about arbitrary dynamical systems. Suppose we are instead interested in the complexodynamics of some system with Hamiltonian $H$ and energy $E$. After some time $t$, unitary evolution takes us from the initial state $\ket{0}$ to  $U(t)\ket{0}$.

 In some cases, this task may be implemented by a classical computer, in the sense discussed above: A time series of steps each involving the application of several orthogonalizing gates. Let $T \equiv n \Delta t$ be the time it takes for the fastest classical computer to implement this task and assume that the energy is constant for simplicity. If the actual computation time of the system, $t$, satisfies 
\begin{align} t\geq T \label{defclass}\end{align}
then one can find a classical computer which does the job just as good or better.  The example \eq{celloyd} is an example of a system in which this can never happen. 

Since $\Delta t$ obeys \eq{lloyd0}, and the actual complexity of the state $U(t)\ket{0}$ must be lower than $n$, the number of gates in the fastest classical computer that produces it,  \eq{defclass} means we obtain a bound in the rate of complexity change,
\begin{align}\dot{\mathcal{C}}\equiv \frac{\Delta \mathcal{C}}{\Delta t}\geq \frac{\pi}{2E},\label{lloyd}\end{align}
 This is the form of Lloyd's bound employed recently in \cite{Brown:2015bva,Brown:2015lvg}. Every system that obeys \eq{defclass} also obeys Lloyd's bound; the converse is not necessarily true\footnote{For instance, depending on our choice of gates in the elementary gate set, there might be some gate $G_i$ in the fastest classical computer which cannot be implemented while saturating the ML bound. In this case, a quantum system might violate \eq{defclass} but still satisfy \eq{lloyd}.}.

So far, the discussion applies to either parallel or serial gate structures.  However, if we assume that time evolution is implementing a series computation then we can write a stronger bound.  More precisely, if we write the time evolution as:
\begin{align}U(t)=\mathbf{T} \prod_i U(t_{i+1},t_i) \end{align}
and assume that $U(t_{i+1},t_{i}) = G_{i}$ is implementing an orthogonalizing gate, then by definition, when the complexity of the circuit increases by one unit the state is already orthogonal.  Since the state complexity must increase more slowly than the circuit complexity, we have 
\begin{align}\dot{\mathcal{C}}\leq \frac{1}{\tau_{orth}}\label{cm}.\end{align}
where  $\tau_{\text{orth.}}$ is the orthogonalizing time of the system.  This is a stronger bound than \eq{lloyd}, applicable only for series circuits.  Furthermore, using (\ref{tml0}), we can write an analogous bound in terms of the time it takes to become `nearly' orthogonal.   We will return to this point in Section \ref{parallel}.  Although there are exceptions, if a system does not obey \eq{cm}, generically we wouldn't expect it to obey \eq{lloyd} either.

\subsection{Final remarks}
To summarize the results of the previous Subsections, we have two classes of subsystems. Any system which takes longer to send $\ket{0}$ to any given state $\ket{\Psi}$ than the best classical computer will obey Lloyd's bound for complexity \eq{lloyd}. We call these systems ``classical''. For some of these, discretized time evolution can be described as a series application of elementary orthogonalizing gates. Then the stronger bound \eq{cm} is satisfied. 

At this point, some important general questions remain.  

\begin{itemize} 
\item {\it{How do we characterize ``classical'' systems and ``classical computers''? In other words, what is the most general class of Hamiltonians such that \eq{lloyd} or \eq{cm} hold, and do field theories with a sensible weakly coupled holographic dual necessarily fall in this class?}}
\end{itemize}
Reference \cite{Brown:2015lvg} proposes that $k$-local Hamiltonians are hopefully enough to saturate Lloyd's bound.  If this is indeed the case, then Lloyd's bound becomes a useful diagnostic tool.  Our emphasis here is instead on the importance of orthogonalizing gates, an assumption which may easily be violated with no apparent pathology.   
 
\begin{itemize}
\item {\it{Is there some other bound that might hold for general systems?}}
\end{itemize}
Clearly, any simple bound will be much looser than ML.  We offer some thoughts on this question in Appendix \ref{somebound}. 

\begin{itemize}
\item {\it{Are Lloyd's bound or \eq{cm} justified for holographic notions of complexity?}}
\end{itemize}  
As we argued above, we a-priori expect that any `smooth' notion of complexity should be based upon simple gates which do not orthogonalize. We will argue next in Section \ref{bhc} that this is indeed the case if the ``complexity = action" conjecture is true; black hole computers are based on simple gates.  There is thus no a-priori reason to expect either Lloyd's bound or \eq{cm} to hold.

\section{Black Hole Computers}
\label{bhc}

Now we turn to our ultimate interest, black holes as quantum computers.  In particular, we would like to know something about the elementary gates being used in a black hole computation.

A priori, we know nothing about the actual gate structure, though we can make some progress by working under the hypothesis that `$\mathcal{C} = \mathcal{A}$'  is true for an appropriate set of gates.  With this assumption, we will argue that holographic gates are simple, in which case the argument for Lloyd's bound given in Section  \ref{lbas} does not apply.  This merely means that holographic systems cannot be thought of as classical computers, but this does not imply any known pathology.   In particular, one does not need to invoke the Weak Gravity Conjecture in order to salvage consistency with computation bounds.  Indeed, {\it{any}} spectral density and {\it{any}} initial wavefunction evolves in a manner consistent with the ML theorem, making it difficult to violate this bound holographically in a trivial way. 

\subsection{Review of `Complexity = Action' }
\begin{figure}[!htb]
\begin{center}
\resizebox{!}{0.4\textwidth}{\begin{tikzpicture}
\node (parab) at (0,0.85) {};
\node[anchor=east,scale=3] (tl) at (-10,5) {$t_L$};
\node[anchor=west,scale=3] (tl) at (10,1) {$t_R$};
\draw[line width=4] (-10,10) -- (-10,-10);
\draw[line width=4] (10,10) -- (10,-10);
\draw[line width=1] (-10,-10) -- (10,10);
\draw[line width=1] (-10,10) -- (10,-10);
\draw[decorate,decoration=zigzag,line width=4] (-10,10) .. controls (-2.5,6.5) and (2.5,6.5) .. (10,10);
\draw[decorate,decoration=zigzag,line width=4] (-10,-10) .. controls (-2.5,-6.5) and (2.5,-6.5) .. (10,-10);
\clip (-10,10) .. controls (-2.5,6.5) and (2.5,6.5) .. (10,10) -- (10,-10) -- (-10,-10) -- cycle;
\filldraw[fill=blue!40!white, draw=blue, line width=3,fill opacity=0.2] (-10,5) -- (2,-7) -- (10,1) -- (-2,13)--cycle;
\draw[decorate,decoration=zigzag,line width=4] (-10,10) .. controls (-2.5,6.5) and (2.5,6.5) .. (10,10);
\end{tikzpicture}}

\end{center}
\caption{Illustration of Wheeler-DeWitt patch in the neutral AdS-Schwarzschild black hole for a pair of times  $(t_{L},t_{R})$. The patch is defined by the intersection of both future and past-directed light rays from both boundary points, and the singularity.}
\label{wdwpatch}
\end{figure}
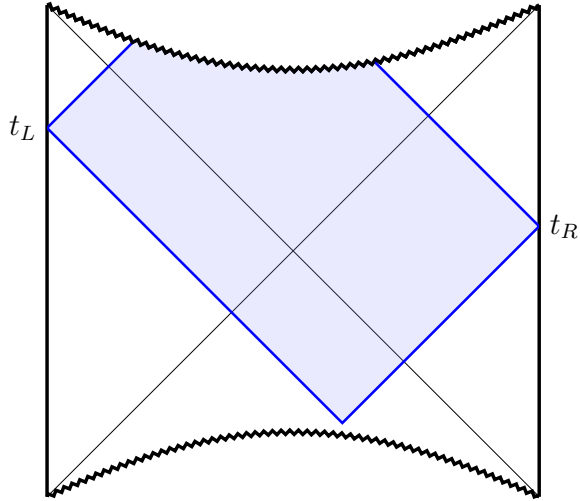

The `Complexity = Action' conjecture relates the action of a certain bulk region in a holographic theory to the complexity of a circuit needed to produce the corresponding boundary state.  More precisely, in the context of two-sided AdS black holes, one picks a pair of times $(t_{L},t_{R})$ on the two boundaries and then defines the Wheeler-deWitt (WdW) patch (Figure \ref{wdwpatch}), consisting of all spacelike paths connecting these two time slices on the boundary.  The claim is then that the action of this WdW patch is equal to the complexity of the right-hand boundary state at $t_{R}$.  The choice of $t_{L}$ is analogous to the choice of a reference state and so this definition suffers from the expected kind of ambiguity.  Obviously, this definition is left-right symmetric; we may also think of this as defining the complexity of the boundary state at the time-slice $t_{L}$, with $t_{R}$ implicitly choosing a reference state.

This and the related `$\mathcal{C} = \mathcal{V}$' conjecture are motivated by the fact that both of these quantities exhibit a linear growth in time long after the system has thermalized.  This alone is already a very powerful observation, since after thermalization all coarse grained observables are static.  We must therefore seek a suitably refined concept to provide a holographic dual to the action growth.  

To make these ideas more concrete, we will consider charged $AdS$ black holes in $D = d+1$ dimensional spacetime:
\begin{eqnarray}
\label{action}
\mathcal{A}&=& \frac{1}{16\pi G} \int_{\mathcal{M}} \sqrt{|g|}\left(\mathcal{R} - 2 \Lambda\right) - \frac{1}{4 e^{2}}\int_{\mathcal{M}} \sqrt{|g|} F_{\mu\nu}F^{\mu\nu} + \frac{1}{8 \pi G} \int_{\partial \mathcal{M}}\sqrt{|h|} K
\end{eqnarray}
where, as usual, $\Lambda$ and the $AdS$ radius are related by:
\be
\frac{(D-1)(D-2)}{L^{2}} = 2 \Lambda
\ee
Solving (\ref{action}) leads us to the solutions:
\beq
ds^{2} &=& - f(r) dt^{2} + \frac{dr^{2}}{f(r)} + r^{2} d\Omega_{D-2}^{2} \\ \nonumber
f(r) &=& 1- \frac{8\pi}{(D-2) \Omega_{D-2}} \frac{2 G M}{r^{D-3}} + \frac{ 8 \pi}{(D-2)(D-3) \Omega_{D-2}^{2}} \frac{G (N e)^{2}}{r^{2(D-3)}}+ \frac{r^{2}}{L^{2}}  \\ \nonumber
F &=& \frac{N e^{2}}{r^{D-2} \Omega_{D-2}} dt \wedge dr
\label{bhmetric}\eeq
\footnote{Note that our conventions differ from \cite{Brown:2015bva} due to a difference in the normalization of the $U(1)$ kinetic terms.  One may relate their conventions to ours by replacing $Q^{2} \rightarrow (e N)^{2}/\left((D-3)\Omega_{D-2}\right)$.} where $N$ is the charge, quantized in units of $e$.  As usual, the temperature is:
\be
\beta = \frac{4 \pi}{f'(r_{s})}
\ee
To illustrate the relevant issues in a slightly simpler setting, we restrict to\footnote{The case of general dimension is discussed in Appendix \ref{gend}.} $D=4$.   In this case, the late time action growth of the Wheeler-deWitt patch for this class of solutions  was computed in \cite{Brown:2015bva, Brown:2015lvg, Cai:2016xho}.   Their result is
\be
\label{dadt}
\frac{d \mathcal{A_{W}}}{dt_{L}} = \frac{(e N)^{2}}{4 \pi}\left(\frac{1}{r_{-}} - \frac{1}{r_{+}}\right)
\ee
The `complexity = action' conjecture states that 
\be
\mathcal{C} = \frac{\mathcal{A_{W}}}{\pi \hbar}
\ee
The precise numerical coefficient is chosen so as to guarantee that large neutral black holes saturate the computation bound\footnote{Thus, we are fixing one overall constant, but the bound is non-trivially satisfied as other parameters are varied, e.g., mass, rotation, and dimension.}.  We may thus translate (\ref{dadt}) into a statement about complexity growth:
\be
\label{dcdt}
\frac{d \mathcal{C}}{dt} = \frac{(e N)^{2}}{4 \pi^{2} \hbar}\left(\frac{1}{r_{-}} - \frac{1}{r_{+}}\right)
\ee
On the other hand, Lloyd's bound applied to this state implies \cite{Brown:2015bva} states.

\be
\label{lloydbh}
\frac{d\mathcal{C}}{dt} \le \frac{2}{\pi \hbar}\left((M- \mu Q) - (M- \mu Q)_{gs}\right)
\ee
where $\mu$ is the chemical potential, $Q = e N$ is the charge, and the subscript $gs$ denotes the ground state.  Of course, the actual ground state depends on the UV completion of the theory.  If, for instance, there are sufficiently light charged particles in the spectrum, then the true ground states could just be a gas of such particles.   For now, we will suppose that the true ground state is just the extremal black hole and explore the consequences of this possibility.  

If the ground state at fixed $\mu$ is given by an extremal black hole, then the RHS of (\ref{lloydbh}) near extremality for a large black hole is
\be
\label{rhs}
RHS \sim \frac{4}{\pi \hbar} \left(M- M_{Q}\right)+\mathcal{O}(M-M_{Q})^{2}
\ee  
(We consider large black holes here since this is the regime in which the worst violation of the complexity bound may be seen.)  Now, on the other hand, the LHS of (\ref{lloydbh}) near extremality is:
\be
\frac{d\mathcal{C}}{dt}\sim \frac{\sqrt{6}}{\pi \hbar} \sqrt{M_{Q}\left(M- M_{Q}\right)}+ \mathcal{O}(M-M_{Q})
\ee
Obviously, for $M$ sufficiently close to $M_Q$ we may violate the bound (\ref{lloydbh}), as first noted in \cite{Brown:2015bva}.  This demands an explanation.  

One possibility is that the black hole simply does not exist in a consistent theory of quantum gravity.  Indeed, this was the proposal of \cite{Brown:2015bva} and is the viewpoint favored by the Weak Gravity Conjecture.  In this scenario, light charged particles would mediate the black hole's decay via Schwinger radiation, thus rescuing the bound (\ref{lloydbh}).  

In contrast, we offer a more mundane explanation. For large black holes, there is no reason to expect \eq{lloydbh} to hold in the first place,  simply  because computation proceeds via simple gates.  According to the discussion in Subsection \ref{rtc},  this means that there is no obstruction to violating the bound \eq{lloyd}.  We are thus in a regime where arbitrarily fast computation is allowed, as described by \cite{2002PhLA..302..291D,Jordan:2017vqh,2017arXiv170105550S}.  In contrast, we do find evidence that for small black holes the bound can be violated while satisfying the orthogonalizing assumption of Lloyd\footnote{The violation for small black holes seems to have been neglected in \cite{Brown:2015bva} and pointed out subsequently in \cite{Cai:2016xho}.}.  To proceed, we will need to understand something about the action of holographic gates on states.  We turn to this next.

\subsection{Margolus-Levitin and `Complexity = Action'}
Our task is now to compute the overlap $\langle \psi | G |\psi\rangle$, where $G$ is an elementary gate and $\psi$ the black hole wave-function. We will first discuss the situation assuming a series gate structure as advocated in \cite{2000Natur.406.1047L} and then mention various caveats that may arise for parallel gates.   Furthermore, we will focus first on the canonical ensemble\footnote{I.e., fixed charge.} since extremal black holes in the grand canonical ensemble are unstable.  We will describe the modifications necessary for the grand canonical ensemble in Section \ref{grand}. For our purposes it will suffice to represent the wavefunction in terms of the thermal field double:
\be
|\psi(t)\rangle = \sum e^{-\frac{1}{2}\beta E_{n} - i E_{n}t} |n\rangle_{L} \times |n\rangle_{R}
\ee
The overlap after evolving by a time $\Delta t$ is
\be
\label{overlap}
\langle \psi(t+ \Delta t) | \psi(t) \rangle = \sum e^{-\beta E_{n} + i E_{n} \Delta t} \sim \int dE\, e^{-(\beta  - i \Delta t)E + S(E)}
\ee
where in the last step we've gone to a continuum approximation and have represented the spectral density as $\rho = e^{S(E)}$.  

We would like to estimate (\ref{overlap}) using the saddle point approximation.  This will allow us to determine the leading result in a large $N$ expansion.  For sufficiently small $\Delta t$ we can further assume that the saddle is effectively the same as for $\Delta t = 0$.  Of course, we will need to examine more carefully the regime of validity of these approximations, which we shall do below.  For now, we simply extract the leading large $N$, small $\Delta t$ result by expanding to quadratic order:
\beq
\label{saddlepoint}
&=&  \int dE\,e^{-\tilde{\beta} (E_{0}+\Delta E) + S_{0} + S'(E_{0}) \Delta E + \frac{1}{2} S''(E_{0}) \Delta E^{2}} \\ \nonumber
&\sim & e^{-\beta F_{0} + i \Delta t E_{0}} e^{\frac{\Delta t^{2}}{2 S''(E_{0})}}
\eeq 
where $\tilde{\beta} \equiv \beta - i \Delta t$.  From this we see that the temporal decay is controlled by:
\be
\label{overlap2}
|\langle \psi(0) | \psi(t)\rangle| = e^{\frac{\Delta t^{2}}{2 S''}}
\ee
Note that $S''(E_{0}) = d\beta/dE$ must be negative for thermodynamic stability\footnote{Near extremal black holes in the canonical ensemble are always thermodynamically stable; they can only decay via Schwinger production.}, so the sign in the exponent makes sense.  Equation (\ref{overlap2}) motivates the introduction of the `coherence time', which, at leading order in $1/N$, is\footnote{Actually, this is a well-known expression in electronic signal theory, but we repeat the derivation here to clarify the underlying assumptions. }
\be \tau_{coh} = \sqrt{-S''(E_{0})} \ee
We will also define a Margolus-Levitin time to facilitate discussion:
\be
\label{TML}
\tau_{ML} = \frac{ \pi}{2 \Delta M}
\ee

Now, we are finally ready to compute the overlap after the application of a single gate\footnote{Again, assuming a series circuit.}.  By the definition of complexity and the conjecture of \cite{Brown:2015bva}, we know that a single gate has operated after a time $\tau_{\text{comp.}}$ has passed, where:
\be
\label{tcomp}
1 = \frac{d\mathcal{C}}{dt} \times \tau_{\text{comp.}}.
\ee
Here $d\mathcal{C}/dt$ is given in (\ref{dcdt}).  Representing this gate abstractly as $G$, the overlap at this time may be written as:
\be
\label{gateoverlap}
|\langle \psi | G |\psi \rangle| = e^{- \frac{1}{2} \left(\frac{\tau_{\text{comp.}}}{\tau_{coh}}\right)^{2}}
\ee

As discussed in Section \ref{lbas}, in order to derive Lloyd's bound, we would need the the gates to nearly orthogonalize.   In contrast, if the gates are simple, there is no bound to be expected.  For the series circuit, we can immediately distinguish between (nearly) orthogonalizing and simple gates simply by comparing the relative size of $\tau_{coh}$ and $\tau_{comp}$.  If $\tau_{comp} \gg \tau_{coh}$ then we have a ``classical computer" where Lloyd's bound holds while if $\tau_{comp} \ll \tau_{coh}$ it does not. 

 In the next Subsection we will show that in fact $\tau_{comp} \ll \tau_{coh}$ for black holes in $AdS_{4}$.   The same statement will be shown in arbitrary dimension in Appendix \ref{gend}.  This firmly establishes that the logical gates are simple assuming the black hole is modeled by a series circuit.   We will offer some comments on the parallel case in Section \ref{parallel}.  Our conclusion in both cases is the same; there is no a priori reason to expect a holographic Lloyd's bound to apply.  

\subsection{Computing $\tau_{coh}$, $\tau_{comp}$ and $\tau_{ML}$}

\begin{figure}
\centerline{\begin{tabular}{cc} \includegraphics[scale = .35]{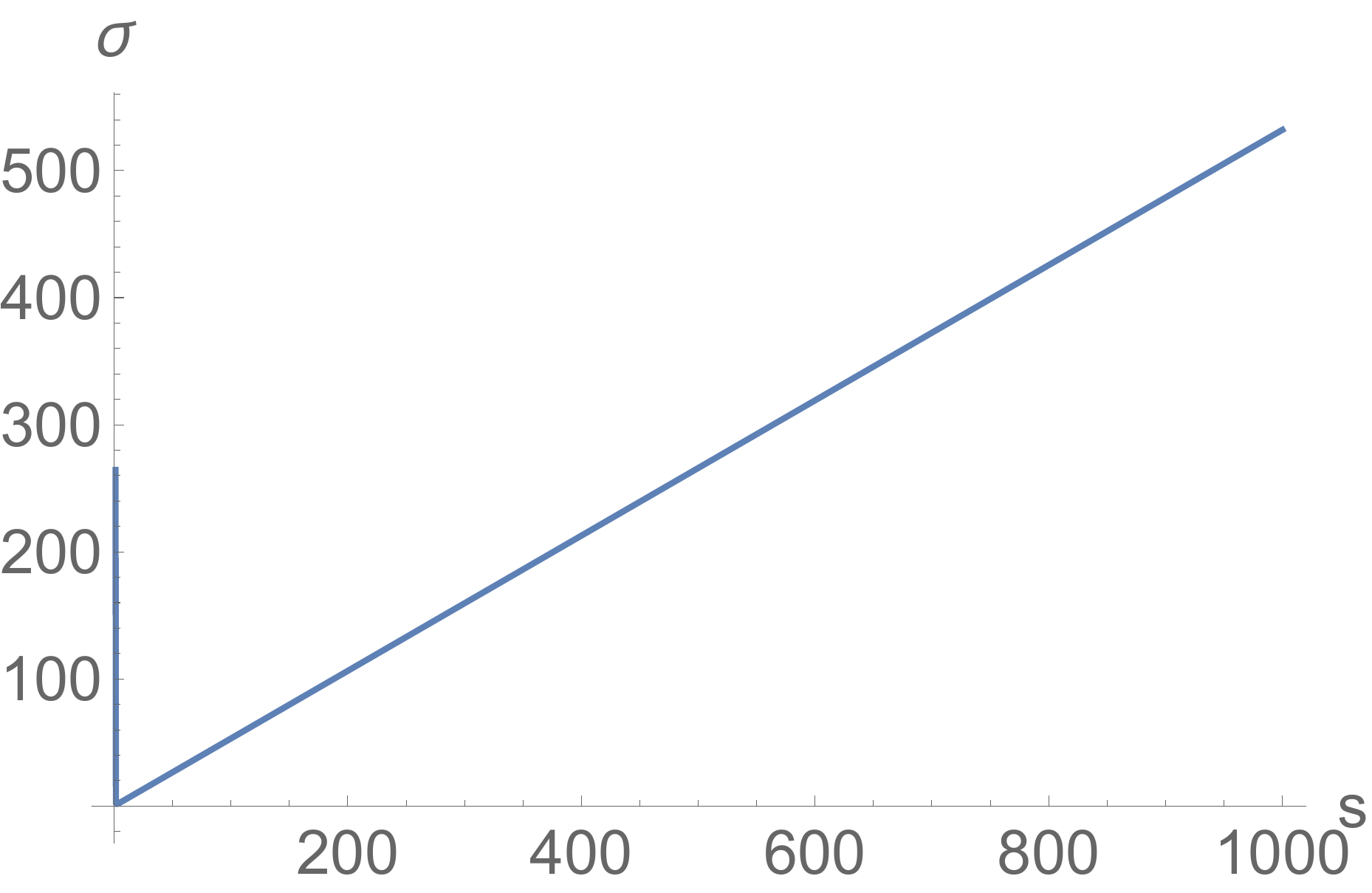} \qquad\,\,&\qquad\,\, \includegraphics[scale = .35]{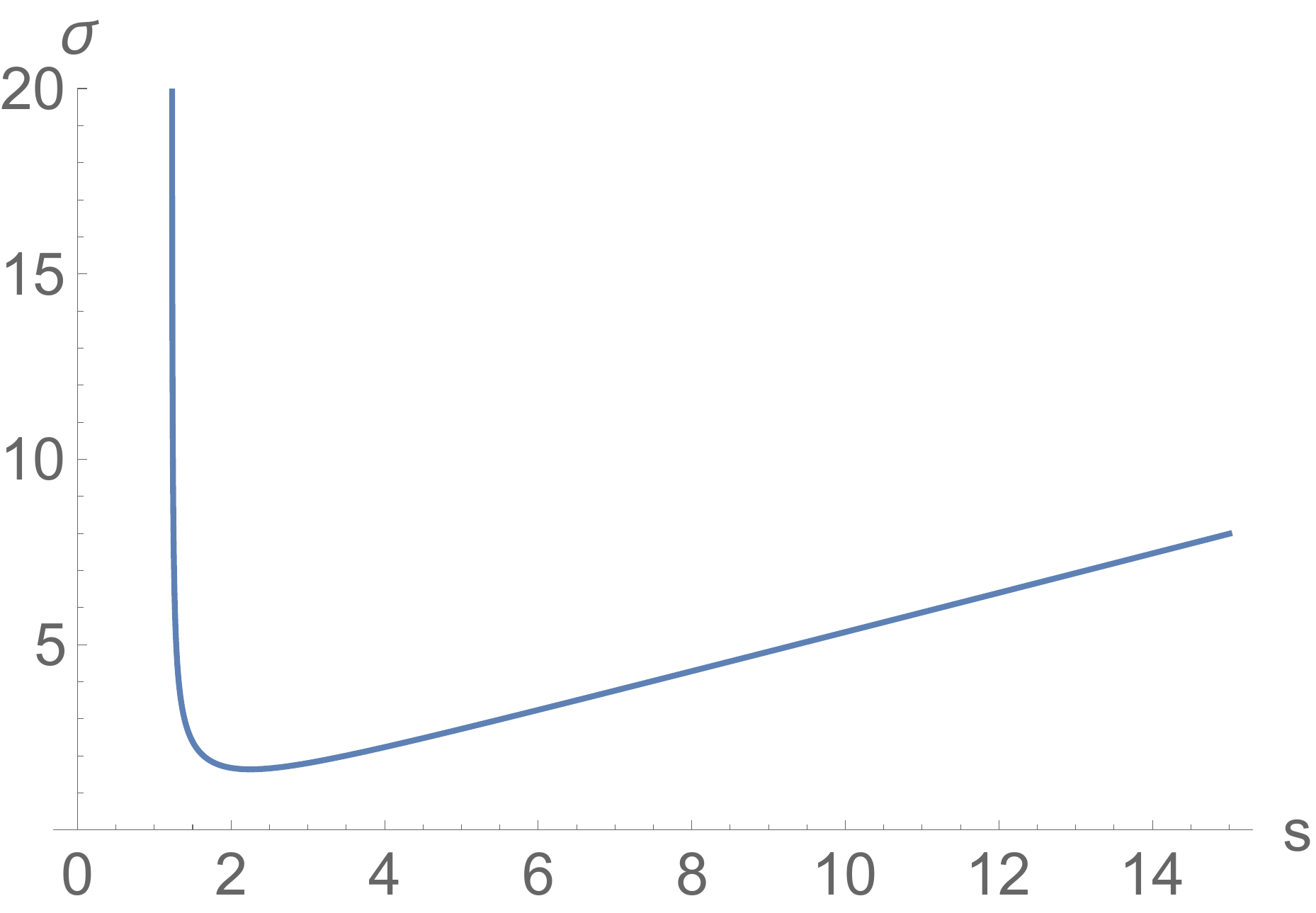} \\ a & b \end{tabular}}
\caption{(a) Plot of $\sigma$ versus $r_{+}/\ell$ for $\tilde{Q} = 10$.  (b)  The same plot, zoomed in to the small $r_{+}/\ell$ region.  One can see that $\sigma$ turns around before ever becoming small.}
\label{sigma}
\end{figure}

Our task is now to compute  $\tau_{coh}$, $\tau_{comp}$ and $\tau_{ML}$, using $S''$,  $d\mathcal{C}/dt$, and $\Delta M$ for the charged black hole.  This is fairly straightforward.  It is convenient to switch to the following rescaled variables
\be
\tilde{Q} = \sqrt{\frac{G}{\ell^{2}}} Q,\qquad \tilde{M} = \frac{G M}{\ell},\qquad s = \frac{r}{\ell}
\ee
in terms of which, the warp-factor becomes
\be
f(s) = 1 + \frac{\tilde{Q}^{2}}{4\pi s^{2}} - \frac{2 \tilde{M}}{s} + s^{2}
\ee
The usual Bekenstein-Hawking entropy is
\be
S(M) = \frac{\pi \ell^{2}}{G} s_{+}^{2}  
\ee
where $s_{+}$ is the largest root of $f(s) = 0$, though the subscript will henceforth be assumed unless otherwise stated.  We want to compute $S''(M) = (G/L)^{2} S''(\tilde{M})$.  Since it is much easier to compute $\tilde{M}(s)$, we write the result in terms of this first:
\beq
S''(M) &=& \frac{128 G \pi^{3} s^{4}\left(3 \tilde{Q}^{2}+ 4\pi s^{2}(3 s^{2}-1)\right)}{\left(\tilde{Q}^{2}- 4\pi s^{2}(1+3 s^{2})\right)^{3}}
\eeq
Finally, we want to compute the following ratio between computation and coherence times
\be
\Sigma \equiv \frac{\tau_{coh}}{\tau_{\text{comp.}}} \equiv \frac{\ell}{\sqrt{G}} \sigma(\tilde{Q},s)
\ee
We are able to factor out an overall $\ell/\sqrt{G} \gg 1$, which is assumed to be large in order for a classical geometric description to be valid.  Thus, the remaining question is whether or not $\sigma(\tilde{Q},s)$ can approach zero in a controlled way.  The answer is no, as illustrated in Figure \ref{sigma}.  Indeed, one can see that $\sigma$ is always at least, $\mathcal{O}(1)$, which means that $\Sigma\gg 1$ in a regime of perturbative control. In general, this implies:
\be
\label{gateoverlap1}
|\langle \psi | G |\psi \rangle| = e^{- \frac{1}{2 \Sigma^{2}} }\sim 1
\ee
Thus, for series circuits, we have shown that the gates relevant for holographic complexity are simple gates and so Lloyd's bound does not apply.    

For completeness, we record the various time-scales both near and far from extremality where it is possible to write simple expressions.

\subsubsection*{Near Extremality}
To compute the relevant timescales near extremality, we expand around the extremal radius $s_{ext}$.  This is
\be
s_{ext}^{2} = \frac{1}{6}\left(\sqrt{1+ x} -1\right)
\ee 
where for simplicity we have introduced $x = \frac{ 3\tilde{Q}^{2}}{\pi}$.  Let us expand, $s_{+}= s_{ext} + \epsilon$.  One finds:
\be
S''(M) = \frac{G \pi\left(\sqrt{1+x}-1\right)^{1/2}\left(\sqrt{1+x} - 1-x\right)}{3 \sqrt{6} (1+x)^{3/2} \epsilon^{3}}
\ee
It is also easy to compute $\Delta M$ by solving $f(s) = 0$ for $M(\tilde{Q},s)$. Then, using (\ref{dcdt}), (\ref{tcomp}) and (\ref{TML}) we get:
\beq
\tau_{\text{comp.}} &=& \frac{G \pi}{\ell\epsilon \sqrt{2+ x + 2 \sqrt{1+x}}}
\\ \nonumber
\tau_{coh} &=&  \left(\frac{G \pi\left(\sqrt{1+x}-1\right)^{1/2}\left( -\sqrt{1+x} +1+x\right)}{3 \sqrt{6} (1+x)^{3/2} \epsilon^{3}}\right)^{1/2} \\ \nonumber
\tau_{ML} &=& \frac{ \pi G }{\sqrt{6} \ell}\left(\frac{x+1}{\sqrt{x+1}-1}\right)^{-1/2} \epsilon^{-2}
\label{qwert}\eeq
Thus, for near extremal black holes we have:
\be
\label{neorder}
\tau_{\text{comp.}} \ll \tau_{coh} \ll \tau_{ML}
\ee
So, indeed, the computation time is parametrically below the ML bound, as first noted in \cite{Brown:2015bva}.  However, at the same time, the gates are simple and so there is no violation of any known computational bound.

\subsubsection{Far from Extremality}
We may repeat this exercise in the opposite limit, $M \gg Q M_{P}$.  In this case, we also want to assume that the black holes are large relative to the $AdS$ scale, otherwise we run into thermal stability issues.  We thus want $\tilde{M} \gg 1$ and $Q$ may be set to zero.  We then find
\be
\tau_{\text{comp.}} = \tau_{ML} = \frac{\pi G}{2 \ell\tilde{M}}, \qquad\qquad \tau_{coh} = \frac{2^{5/6}\sqrt{\pi} \sqrt{G}}{3 \tilde{M}^{2/3}} 
\ee
and so
\be
\label{feorder}
\tau_{comp} = \tau_{ML} \ll \tau_{coh} 
\ee
Again, as noted \cite{Brown:2015bva}, the ML bound is apparently saturated for large black holes\footnote{Though this is partially by construction}.  However, in light of the fact that again $\tau_{\text{comp.}} \ll \tau_{coh}$,  there is no rigorous reason to believe that known computation bounds are justified in the first place.

\subsection{ML versus decay?} \label{mlvsdecay}
One might wonder whether or not we may simply bypass Lloyd's bound and look for some conflict with Margolus-Levitin directly?  In particular, is it a problem if the coherence time is much less than the ML time as indicated in (\ref{neorder}) for near-extremal black holes?  As we argued in Section \ref{MLS}, the decay time must always be of order the actual orthogonality time and so an ML bound should apply.  What is going wrong?

To address this issue, we must return to inspect the validity of the approximations taken, in particular, the large $N$ saddle point.  In equation (\ref{saddlepoint}) we expanded $S(E)$ to quadratic order, assuming that the higher order terms are subleading.  In particular, we must assume $|S''(E)| \gg S''(E) \Delta E$. Moreover, the rms fluctuation of $\Delta E$ is $\sim 1/\sqrt{-S''}$.  Putting this together, in order to trust the saddle point approximation we need:
\be
C \equiv \frac{(-S'')^{3/2}}{S'''} \gg 1
\ee
Typically, this ratio is large since it scales like $S^{1/2}$.  However, near extremality we have
\be
\label{spc}
C = \frac{(-S'')^{3/2}}{S'''}  \sim \begin{cases}   \epsilon^{1/2} x^{1/8} &\,\,\, {\text{\bf{Large BH's}}} \\ \epsilon^{1/2} x^{1/4} & \,\,\,{\text{\bf{Small BH's}}} \end{cases}
\ee
In other words, sufficiently close to extremality, ($\epsilon \ll 1$) our control parameter becomes small and the saddle point breaks down.  We can see the issue more directly by writing the ratio of near-extremal timescales in terms of $C$:
\be
\frac{\tau_{coh}}{\tau_{ML}} \sim \frac{\ell}{\sqrt{G}}\,C
\ee
Thus, whenever the saddle point is valid $\tau_{coh}/\tau_{ML}$ is automatically large, in agreement with the ML theorem.  This was inevitable, as the ML theorem applies to any wavefunction evolving via a Hamiltonian.  The apparent discrepancy in equation (\ref{neorder}) is a special case of $[\lim_{N\rightarrow \infty} \cdot, \lim_{T\rightarrow 0} \cdot ] \ne 0$.

\subsection{Grand Canonical Ensemble}
\label{grand}
In the grand canonical ensemble we must allow for the energy and charge to fluctuate, which means that the partition function looks like
\be
Z = \int dE dQ e^{- \beta(E- \mu Q) + S }
\ee
Since there are more allowed fluctuations it is intuitively clear that the coherence time will decrease. Thus, one may wonder whether it is possible to reach the situation $\tau_{coh} \ll \tau_{comp}$, thus justifying Lloyd's bound.  Here we show that this does not happen near extremality,  but rather, on the boundary of stability between $AdS$ and black holes solutions.  In this case, one indeed finds that the assumptions leading to Lloyd's bound do not hold.  

Following again the derivation in (\ref{saddlepoint}), one finds that the coherence time is now:
\be
\tau_{coh} = \left(-v^{i} (\partial_{i}\partial_{j} S)^{-1} v^{j}\right)^{-1/2} = \left(-\partial_{\beta}^{2}( \beta G(\beta,\mu))\right)^{-1/2}
\ee
where $v^{i}$ is shorthand for $\vec{v} = (1, -\mu)$ and $G(\beta,\mu)$ is the usual Gibbs free energy:
\be
G(\beta,\mu) = \frac{\ell s}{16 G}\left(4 - \pi G \mu^{2} - 4 s^{2}\right)
\ee
where the equilibrium relations are
\be
\beta = \frac{16 \pi \ell s}{4 - \pi G \mu^{2} + 12 s^{2}} \qquad\qquad \mu = \frac{Q}{4\pi \ell s}\label{er}
\ee
Now it is straightforward to compute the coherence time.  For large black holes the situation is exactly the same as we found before; namely, the coherence time near extremality goes like
\be
\tau_{coh} \sim \frac{\sqrt{G}}{\epsilon^{3/2}}
\ee
where $\epsilon = s-s_{ext}$.  Again, we find that the assumptions of Lloyd's bound do not hold in this regime.  

\begin{figure}
\centerline{ \includegraphics[scale = .5]{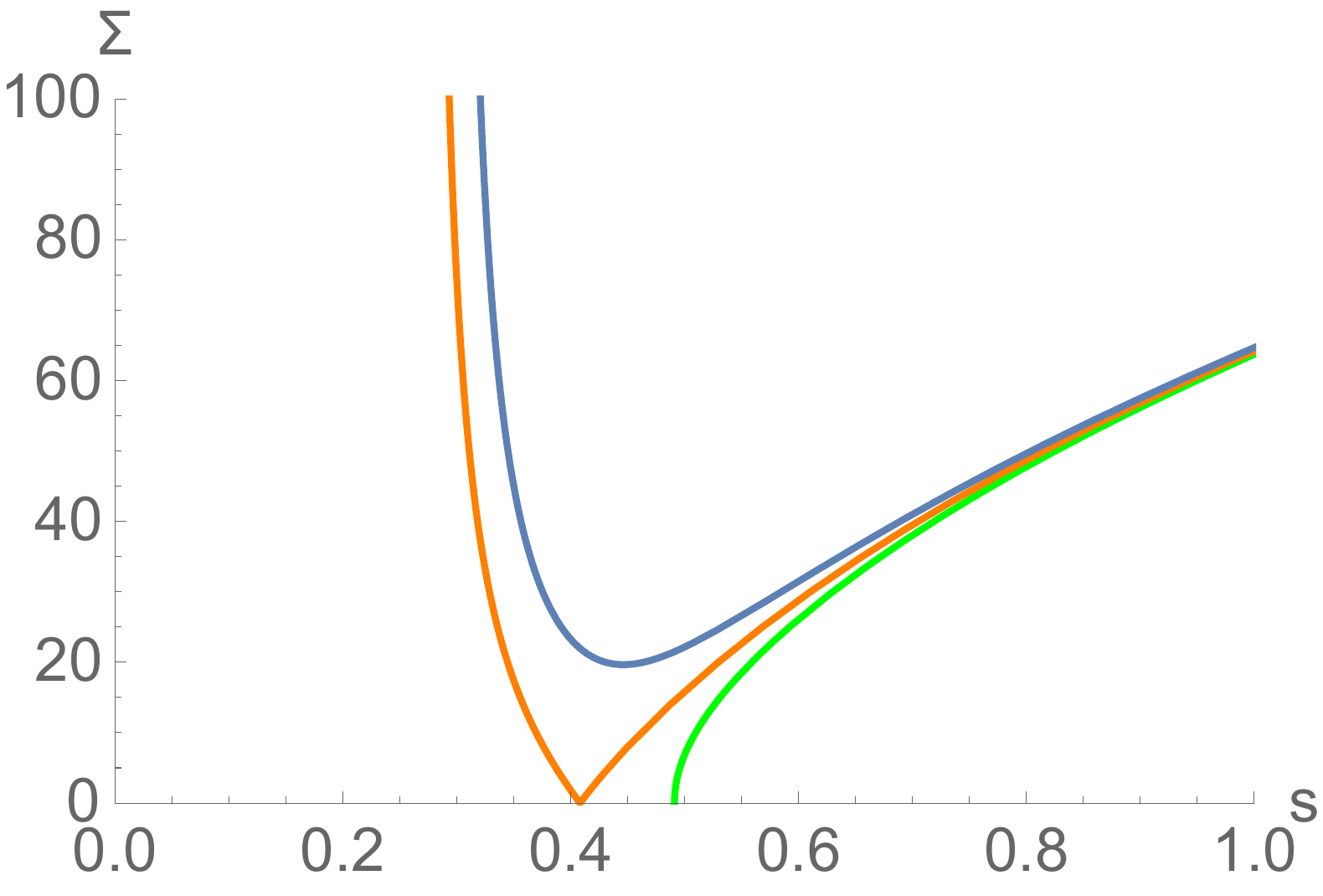} }
\caption{$\Sigma$ as a function of the black hole radius $s$ for fixed charge $ =1.2, 1$ and $.8$ depicted in blue, orange and green, respectively.  Here we have chosen $\ell/\sqrt{G} = 100$.}
\label{grandsigma}
\end{figure}

The situation with small black holes is slightly more subtle. The above discussion applies for near-extremal black holes, which is good enough for large chemical potential, but as pointed our, for instance, in \cite{Chamblin:1999tk},  there are regions of the $(\mu,T)$ plane in which one cannot continuously approach the extremal limit for small charged black holes. This is just the charged generalization of the Hawking Page transition, marking the boundary of non-perturbative stability.  Inside the unstable region there is also a window of perturbative stability, which will be of interest here.  In our notation, for $\mu < 2/\sqrt{\pi G}$, there is a minimal critical temperature $T_{min}(\mu)$ below which a perturbatively stable solution ceases to exist and one only has the $AdS$ vacua at finite chemical potential.  Since the black hole becomes perturbatively unstable at this temperature, it would stand to reason that the coherence time goes to zero as well.  Indeed, this is the case, as illustrated in Figure \ref{grandsigma}, where we see that $\Sigma = \tau_{coh}/\tau_{comp}\rightarrow 0$ precisely at this threshold.   This plot illustrates the behavior of $\Sigma$ along a curve of constant charge (defined by the second equation in \eq{er})  in the $(\mu,T)$ plane.

\begin{figure}
\centerline{ \includegraphics[scale = .4]{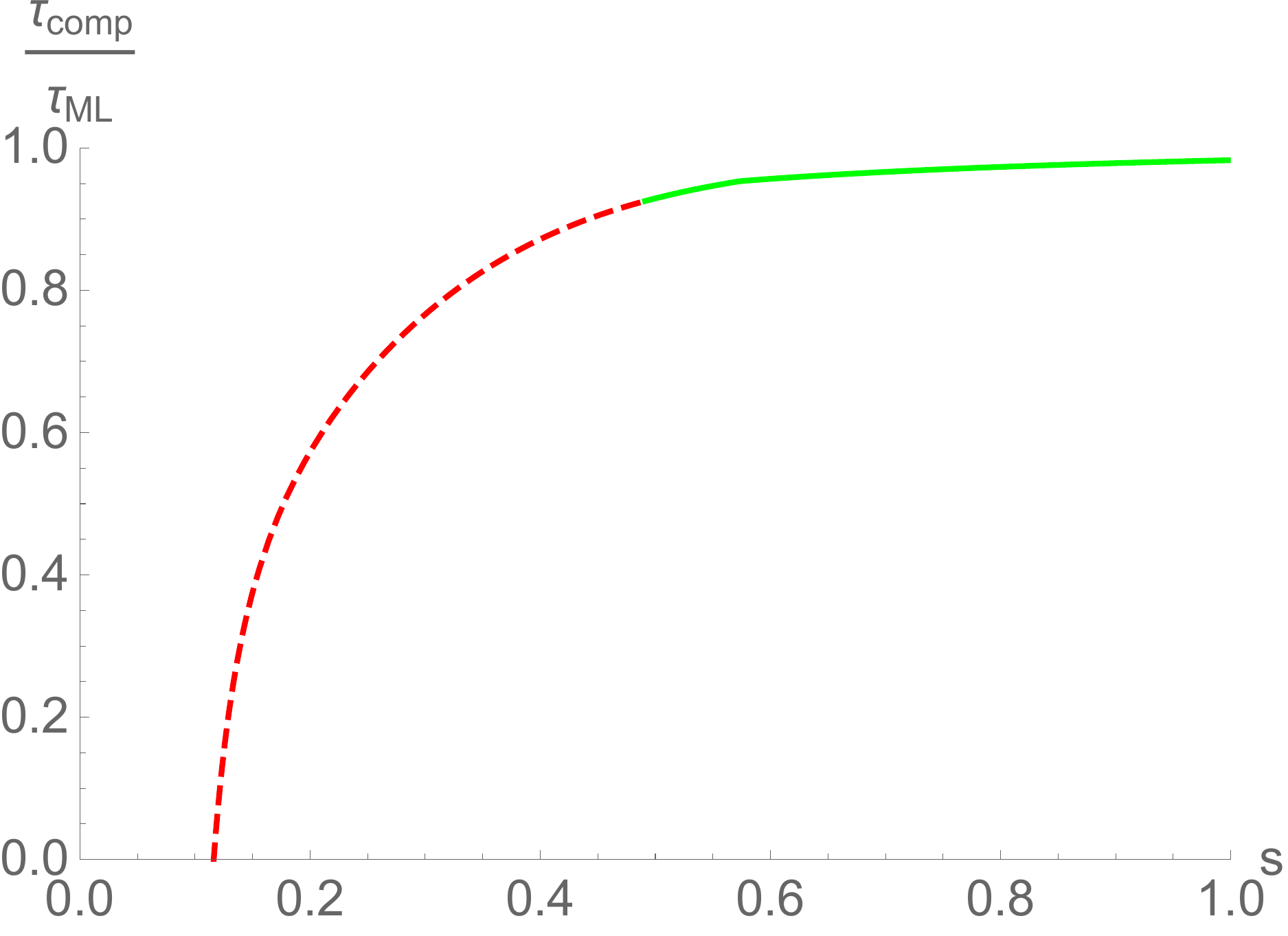} }
\caption{A violation of Lloyd's bound at $x=.8$.  The assumptions of Lloyd's bound are satisfied at the transition region between stable (blue) and unstable (red) regions.}
\label{rimp}
\end{figure}

It is also interesting to consider the ratio $\tau_{comp}/\tau_{ML}$ near the phase transition where Lloyd's bound should presumably hold. Since we are now in the grand canonical ensemble, it is natural to define a new $\tau_{ML}$, following the conjecture of \cite{Brown:2015bva}, as
\be
\label{tmlgrand}
\frac{1}{\tau_{ML}} = \frac{2}{\pi}\left((M- \mu Q) - (M- \mu Q)_{gs}\right)
\ee
where $(M- \mu Q)_{gs}$ denotes the ground state energy at fixed chemical potential.   In Figure \ref{rimp} one sees that the ratio $\tau_{comp}/\tau_{ML}$ can be less than one, {\it{even}} in the region where $\Sigma \ll1$.  This is the essence of the argument presented in \cite{Brown:2015bva}, which provides some support for the Weak Gravity Conjecture.  

We note, however, that to leading order in the large $\ell/\sqrt{G}$ limit, any non-zero value of $\Sigma$ in Figure \ref{grandsigma} is actually scaled to infinity.  Thus, within the sugra approximation Lloyd's bound is never justified, except precisely at the threshold $\tau_{coh}=0$.  Our results indicate that there is indeed a violation at this point.  However, one should take this with a grain of salt since one loop effects are potentially large near the threshold and could alter the classical result in a manner analogous to the situation described with the canonical ensemble in Subsection \ref{mlvsdecay}. 

\subsection{Parallel Gates}
\label{parallel}
So far in our discussion we have assumed that black holes are series circuits as has been argued on general grounds \cite{2000Natur.406.1047L}.  In particular, the assumption was used in (\ref{gateoverlap}) where after a time $\Delta t$ we arrive at a state $G(\Delta t)\ket{\Psi}$.  Let us now briefly consider a parallel computer.  In contrast to (\ref{gateoverlap}), in the same time period a parallel processor would evolve the state as
\be
\ket{\psi} \rightarrow G_{1}(\Delta t)G_{2}(\Delta t)... G_{n}(\Delta t) \ket{\psi}
\ee  
where again we must assume that each $G_{i}(\Delta t)=G_{i}(n \tau_{\text{comp.}})$ orthogonalizes the state $\ket{0}$ and the $G_{i}$ must commute in order for the proof of Lloyd's bound to hold.

It is not trivial to satisfy these conditions on the $G_{i}$'s, particularly if we require these conditions to be true for a family of states. One particular structure that will work is if the Hilbert space, $H$, is a tensor product $H = H_{1}\otimes H_{2} \otimes... \otimes H_{n}$, where each $G_{i}$ acts on a particular $H_{i}$. Such a gate structure could be motivated by locality, though our only concern at the moment is understanding conditions under which Lloyd's bound would necessarily hold.  With the tensor product assumption, one immediately obtains $\Delta t > \tau_{coh}$.  Thus, to show that the gates {\it{cannot}} be orthogonalizing, we would need

\be 
\label{killL}
\Sigma = \frac{\tau_{coh}}{\tau_{\text{comp.}}} > n
\ee
For near extremal black holes we in the canonical ensemble have seen $\Sigma \sim \frac{\ell}{\sqrt{G}}\epsilon^{-1/2}$.  Thus, for any fixed $n$ it appears that (\ref{killL}) is indeed satisfied close enough to extremality or for large enough $\ell$.  However, when we take into account the condition (\ref{spc}) for the validity of the saddle point, we find that (\ref{killL}) becomes\footnote{The bound below comes from looking at large black holes.  For small black holes, the $\epsilon$ expansion breaks down when the saddle point is valid.}

\be
\label{ncond}
n \lesssim  \frac{\ell}{\sqrt{G}} \left(\frac{Q \sqrt{G}}{\ell}\right)^{1/4} 
\ee

In other words, with the tensor product structure assumed above, the gates are simple unless the degree of parallelization exceeds the bounds given in (\ref{ncond}).  It seems plausible that improved bounds could be obtained with a more careful analysis.  For reference's sake, \cite{Brown:2015lvg} claims that each gate must increase the complexity by roughly the central charge, which, we could interpret as a claim of parallelization being of order `$c$'.  A further comparison between this requirement and the above bound seems warranted.  

The above discussion shows that, with a tensor product structure, Lloyd's bound is not excluded, even for large black holes, provided the circuit is sufficiently parallel. One may still wonder if there are even weaker sets of assumptions under which the bound might be viable.  For instance, one could imagine including auxiliary factors in the tensor product, i.e., $H = H_{1}\otimes H_{2} \otimes... \otimes H_{n}\otimes A_{1}\otimes A_{2}...$. We will regard the factors $H_i$ as ``observable'', and the $A_i$ will be ``unobservable'', and choose a set of gates that acts simultaneously in the observable and unobservable sectors.  By requiring the gates to orthogonalize only in the unobservable factors, $A_{i}$, the action of the elementary gates in the physical sector is essentially unconstrained\footnote{There are, however, implicit restrictions coming from the requirement that each gate has positive energy in the full system.} and one may hope to sneakily evade our conclusions.  In other words, this is an example of a gate structure that leads to Lloyd's bound, but such that the observable dynamics is only very mildly constrained.  

However, before becoming too optimistic, we warn that requiring the system to orthogonalize the $A_i$ necessarily forces a slowdown of the computation, such that Lloyd's bound  still holds, but can never be saturated.   Moreover, we must introduce a much enlarged Hilbert space and choose a peculiar set of finely-tuned gates. 
  
At this moment we have not excluded the possibility that consistent, reasonable gate choices may be made, but we hope this discussion serves to illustrate the burden that might be faced in showing that the necessary assumptions for Lloyd's bound have a chance of holding in the regime of large black holes.  At the very least, one must demand that black holes computers are highly parallel, and may be constructed from orthogonalizing gates rather than simple ones.  

\section{Future Directions and Conclusions}
\label{conclusions}

The main goal of this paper has been to pinpoint the assumptions underlying Lloyd's bound in relation to holographic complexity. One of our motivations to do this was the connection to the Weak Gravity Conjecture: References \cite{Brown:2015bva,Brown:2015lvg} showed that, assuming $\mathcal{C}=\mathcal{A}$, every reasonable AdS black hole satisfied the bound, the one exception being the large, charged, near-extremal RN-AdS black hole. These are precisely the black holes that should not exist in any AdS theory of gravity which satisfies the WGC. So, if it could be argued somehow that Lloyd's bound for complexity was an actual sharp bound that should always be obeyed by reasonable holographic systems, this could be turned into an argument for the WGC.  In other words, failure to saturate Lloyd's bound would be a ``diagnostic tool'' for IR theories which do not have a sensible UV completion and thus lie in the Swampland.

Instead, we found what is a well-known result in the quantum information literature \cite{2002PhLA..302..291D,Jordan:2017vqh,2017arXiv170105550S} - namely that a generic quantum system will violate Lloyd's bound. Systems that obey the bound can be replaced by a classical computer working at the same energy.  Systems that violate the bound must be represented by a quantum computer but they are not inconsistent.

Of course, a holographic theory is not just any generic quantum system, so perhaps there is a stronger set of properties which mean that Lloyd's, or something like it, must always be satisfied. We analyzed a simple set of assumptions that implies Lloyd's: That time evolution can be described as the series application of orthogonalizing gates, just like in an ordinary classical computer. This set of assumptions implies a stronger bound \eq{cm}, which makes analysis particularly easy. Our results on the different AdS black holes are summarized in the following table, assuming a serial gate structure:

\medskip

\begin{center}
\begin{tabular}{|c|c|c|}\hline \textbf{Black hole} &\vtop{\hbox{\strut \textbf{Does it satisfy the assumptions}}\hbox{\strut \textbf{of Lloyd's bound?}} } & \vtop{\hbox{\strut \textbf{Does it actually}}\hbox{\strut \textbf{obey the bound?}} } \\\hline
Schwarzschild& No & Yes \\\hline
Small RN & No & Yes\\\hline
Large RN & No & No\\\hline
\end{tabular}
\end{center}

\medskip

The fact that the first column is ``No'' in all three cases means that time evolution proceeds via simple gates - assuming it is serial. Had the third row been ``Yes, No'', we would have had a hierarchy
\begin{align*}\tau_{coh}<\tau_{\text{comp.}}< \tau_{ML},\end{align*}
which would have been in contradiction with the Margolus-Levitin theorem. This, in turn, would have signaled a real pathology in the system. 

Since this is not the case, we cannot offer strong support for the WGC. Rather, our results mean that, at least from the perspective of the complexodynamics of  black holes, there is nothing obviously wrong with the large near extremal RN-AdS - it is just a generic simple system (meaning that evolution proceeds via simple gates) that doesn't satisfy Lloyd's bound. From this point of view, it is the fact that the other two examples do saturate the bound that is somewhat surprising at first.  There may very well be some deep reason for this, other than dimensional analysis, but if so, it must be something very different from the usual justification for Lloyd's bound. To our knowledge, such an alternative has not yet been discussed in the literature.

We should also mention that all the above results are valid for both the canonical and the grand canonical ensembles, with one caveat - along the coexistence curve in the $(\mu,T)$ plane in the grand canonical ensemble we do get a seemingly pathological violation of Lloyd's bound - but always in a regime in which one should be wary of our approximations. 

There are many interesting open questions. Perhaps the most outstanding is whether there is any other set of reasonable assumptions that could lead to a proof of Lloyd's bound for semiclassical states in holographic systems. If this is not the case, is there any other rigorous complexification bound that might be used as an IR diagnostic?  Finally, one should address the proper interpretation of the failure of Lloyd's bound along the coexistence curve where the strongest case for the bound's validity may be made.  We hope to return to these issues in the near future.
 
  \subsection*{Acknowledgements}
   We thank Shira Chapman, Ben Freivogel, Ro Johnson, Sagar Lokhande, Gary Shiu and Pablo Soler  for valuable discussions and comments. MM is supported by a postdoctoral fellowship by ITF, Utrecht University.

\noindent\begin{appendices}

\section{General bounds on the rate of complexity change}
\label{somebound}

It is possible to derive a general bound on (a regularized version of) the rate of complexity change, which works even for quantum computers, but it is much weaker than \eq{lloyd}. 

Fix $\epsilon$ and the dimension of the Hilbert space, $D$. The complexity of any state attains a maximum $n_{\text{max}}(\epsilon, D)$, precisely when there are enough points to cover all of the projective Hilbert space with balls of radius $\epsilon$ center at the points in $\mathcal{F}$. The plots in Figure \ref{fig1} illustrate the fact that the complexity of a typical state will be close to $n_{\text{max}}$ - states of low complexity are a tiny fraction of this. Under this assumption, we can assume that the projective Hilbert space is covered by roughly $2^{n_{\text{max}}(\epsilon, D)}$ balls  of radius $\epsilon$. Using the volume element associated to the Fubini distance, the volume of a ball of radius $R$ in $D$-dimensional space is
\begin{align}V_D(R)=\frac{\Omega_{2D-3}}{2^{2D-3}}\int_0^Rdr\, \vert \sin( 2r)\vert ^{2D-3} ,\end{align}
from which we get the estimate
\begin{align}n_{\text{max}}(\epsilon, D)\approx\log\left(\frac{V_D(\pi/2)}{V_D(\epsilon)}\right)=\log\left(\frac{\sqrt{\pi}}{2}\frac{\Gamma(d)}{\Gamma(d-1/2)}\epsilon^{2(1-D)}\right)\approx -2D\log(\epsilon).\end{align}
This estimate should be good for large enough $\epsilon$ and $D$. 

Now, imagine we are at some state in the Hilbert space with low complexity $\mathcal{C}_0$. Time evolution will move the state around; by the time it has moved a distance $\Delta x\sim 2\epsilon$ away from its original location, it will for sure have left the ball for another, and its complexity will have changed to $\mathcal{C}_1$. Clearly, we have $\Delta\mathcal{C}\leq n_{\text{max}}(\epsilon, D)$.

We can also easily estimate the time it took the state to move $\Delta x$, if $\epsilon$ is very small, since for small times the distance $d(\ket{\psi(0)},\ket{\psi(t)})$ goes as 
\begin{align}d(\ket{\psi(0)},\ket{\psi(\Delta t)})\approx\Delta E \Delta t.\end{align}
Here, $\Delta E\equiv\sqrt{\langle H^2\rangle - \langle H\rangle^2}$ is the standard energy dispersion of the state. Putting everything together, we get 
\begin{align}\frac{\Delta \mathcal{C}}{\Delta t}\leq \frac{n_{\text{max}}(\epsilon, D)}{\epsilon} \Delta E.\label{bound0}\end{align}
This is a hard bound on the rate of change of complexity, which always works, but it is not very helpful, since  $n_{\text{max}}(\epsilon, D)/\epsilon$ diverges for $\epsilon\rightarrow0$. 
To turn \eq{bound0} into something useful, we need to work with some regularized complexity which removes the $\epsilon$ divergence. Thus, we define $\mathcal{C}_R= \frac{\epsilon}{-\log(\epsilon)}\mathcal{C}$. Then, we have
\begin{align}\frac{\Delta \mathcal{C}_R}{\Delta t}\leq 2D \Delta E.\label{bound1}\end{align}
Unlike Lloyd's bound \eq{lloyd}, \eq{bound1} applies at any time, and not just after a time step. This bound is still pretty lax, since $D$ here is the dimension of the part of the total Hilbert space in which the state moves. Suppose now that the state is actually pretty close to an energy eigenstate, so that $\Delta E\ll E$. The dimension of the Hilbert space where the particle lives is $D\approx\exp(S(E))\Delta E$. The above becomes
\begin{align}\frac{\Delta \mathcal{C}_R}{\Delta t}\lesssim \exp(S(E)) (\Delta E)^2,\end{align}
which is still a pretty large bound. Perhaps it looks nicer in terms of the regularized logarithmic complexity $\log \mathcal{C}_R$, in terms of which we have
\begin{align}\frac{\Delta \log \mathcal{C}_R}{\Delta t}\lesssim S(E) (\Delta E)^2.\end{align}
 As an illustration, suppose furthermore that the state we are considering is e.g. a black hole, i.e. it is some approximately thermal state in which
\begin{align}(\Delta E)^2= \frac{dE}{d\beta}.\end{align}
Then, we have
\begin{align}\frac{\Delta \log \mathcal{C}_R}{\Delta t}\lesssim S(E) \frac{dE}{d\beta}.\label{bound2}\end{align}
For a Schwarzschild black hole, the right hand side is $\propto E^2$, so this bound is much weaker than \eq{lloyd}. A bound such as \eq{bound1} or \eq{bound2}, while pretty general, is probably not very useful, since it is saturated by states whose complexity becomes maximal as soon as possible. Furthermore, it is a bound on a  particular regularized version of complexity; it is conceivable that different regularizations lead to different bounds, though we expect the underlying reasoning to always hold. 

\section{Black Holes in General Dimensions}
\label{gend}

In this Appendix we generalize the results in Section \ref{bhc} to arbitrary dimension. 
\subsection{Fixed charge ensemble}
The orthogonalization time is then defined just as in the main text,
\begin{align}\tau_{\text{orth.}}=\frac{1}{\sqrt{-(\beta F)''}}=\frac{1}{\sqrt{ -E'(\beta)}}.\end{align}
where primes denote derivatives with respect to $\beta$, and in the last equality we used that $(\beta F)'=E$. This is only meaningful if $\tau_{\text{orth.}}\ll\beta$. 

Now we should evaluate the above for charged RN-AdS black holes. Refs. \cite{Chamblin:1999tk,Chamblin:1999hg} contain the relevant expressions. For a black hole of radius $r$, in $d+1$ dimensions, we have 
\begin{align}M(r)=\frac{(d-1)\Omega_{d-1}}{16\pi G}\left( r^{d-2}+\frac{q^2}{r^{d-2}}+\frac{r^d}{\ell^2}\right).\label{mmm}\end{align}

$q^2$ is just the coefficient of the second to last term in \eq{bhmetric}. On the other hand, the radius $r$ is related to the temperature and $q$ via
\begin{align}\beta=\frac{4\pi \ell^2 r^{2d-3}}{dr^{2d-2}+(d-2)\ell^2r^{2d-4}-(d-2)q^2\ell^2}.\label{betis}\end{align}
From these two it is straightforward to compute (introducing $s\equiv r/\ell$, and $\tilde{q}=q/\ell^{d-2})$
\begin{align}\tau_{\text{orth.}}^2=-\frac{dE}{dr}\left(\frac{d\beta}{dr}\right)^{-1}=\frac{G}{\ell^{d-3}}\frac{64 \pi ^2  s^{3 d+3} \left((d-2) (2 d-3) s^4 \tilde{q}^2+\left(d \left(s^2-1\right)+2\right) s^{2 d}\right)}{(d-1) \Omega _{d-1} \left(s^{2 d} \left(d s^2+d-2\right)-(d-2) s^4 \tilde{q}^2\right)^3}.\end{align}

\begin{figure}
\centering
\includegraphics[scale = .65]{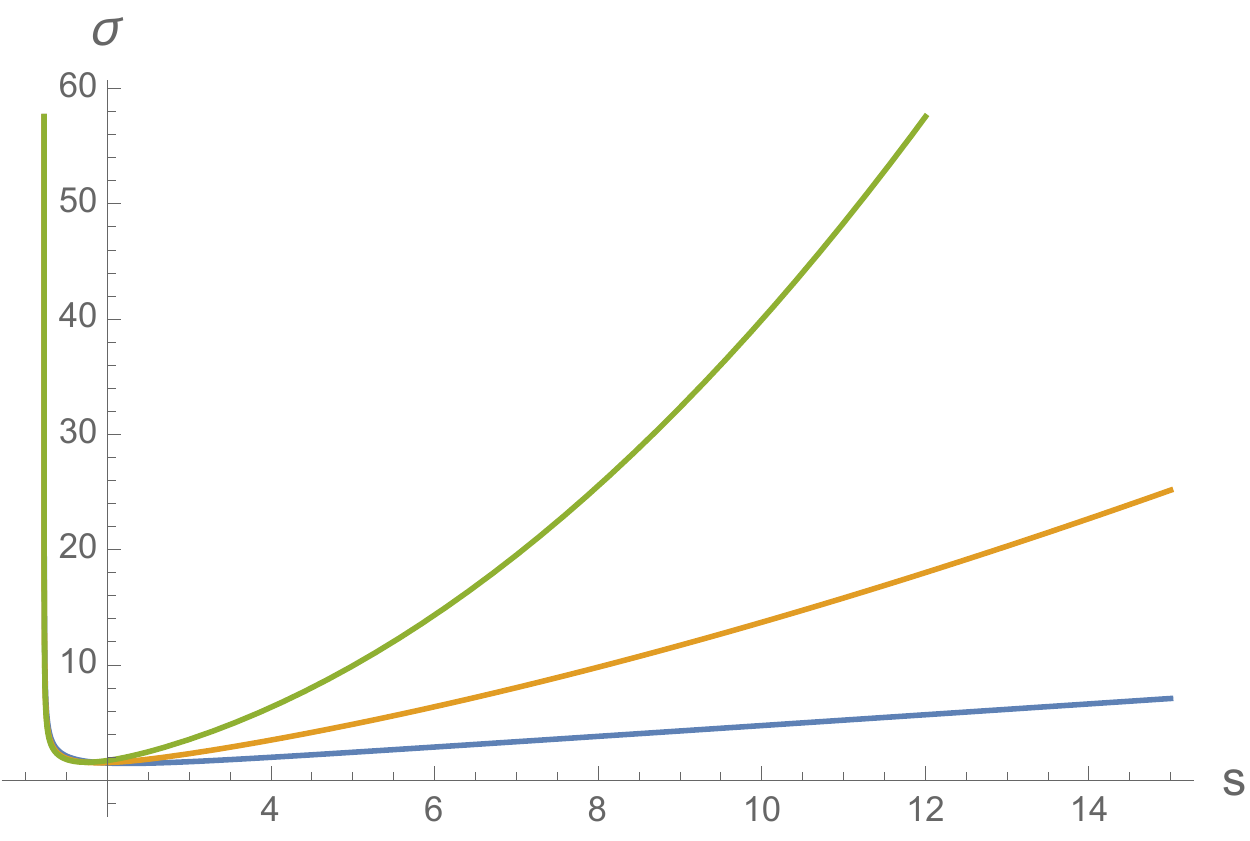}
\caption{ Plot of $\sigma$ versus $s$ for $y = \sqrt{3/2}$, for spacetime dimension $d+1=4,5,6$ (blue, orange, and green lines, respectively).  The results are qualitatively the same for every $d$.}
\label{sigmad}
\end{figure}

According to \cite{Cai:2016xho} the complexification timescale $\tau_{\text{comp.}}$ given by 
\begin{align}\tau_{\text{comp.}}=\frac{G}{\ell^{d-1}}\frac{1}{\Omega_{d-1}(d-1)}\frac{1}{s^{d-2}-x_-^{d-2}+s^{d}-s_-^{d}},\end{align}
where $x_-$ is the inner horizon radius (as a function of $x$ and $\tilde{q}$) 
From this, we get the $\sigma$ function in $d$ dimensions:
\begin{align}\frac{\tau_{\text{orth.}}}{\tau_{\text{comp.}}}\equiv\frac{\ell^{\frac{d-1}{2}}}{\sqrt{G}}\sigma_d(\tilde{q},s). \end{align}
Figure \ref{sigmad} shows plots for spacetime dimension $d+1=4,5,6$, illustrating that the qualitative structure is similar to the four-dimensional case.
We also want to understand the near-extremality behaviour. Expanding again $x=y+\epsilon$, where $y$ is the extremal radius, we get
\begin{align}\tau_{\text{orth.}}&\approx\frac{\sqrt{G}}{\ell^{\frac{d-3}{2}}}\frac{4 \pi  \sqrt{\frac{ y^{6-d}}{(d-1) \Omega  \left((d-1) d y^2+(d-2)^2\right)^2}}}{\epsilon ^{3/2}}+\mathcal{O}(\epsilon^{-1/2}),\\ \tau_{\text{ML}}&\approx \frac{G}{\ell^{d-1}}\frac{8 \pi ^2  (d-2)  y^{4-d}}{(d-2) (d-1) \Omega_{d-1}   \left((d-1) d y^2+(d-2)^2\right)}\frac{1}{ \epsilon ^2}+\mathcal{O}(\epsilon^{-1}).\end{align}
These agree with \eq{qwert} when $d=3$. To get the approximate expression for $\tau_{\text{comp.}}$, we need to work a little bit harder, since there is no closed expression for $x_-$. We will parametrize $s=y+\epsilon$, $s_-=y+\alpha\epsilon$.  In terms of these, we have
\begin{align}\tau_{\text{comp.}}=\frac{G}{\ell^{d-1}}\frac{8\pi  (d-2) y^{6-2 d}}{ \left( (d-1) \Omega  \left(d y^2+d-2\right)\right)}.\frac{1}{2 \alpha -\alpha  d+d-2}\frac{1}{\epsilon}\end{align}
Our task is determining $\alpha$, to first order in $\epsilon$. Let $P(z)=(z-x)(z-x_-)Q(z)$ be the defining polynomial. We have
\begin{align}P(y)=\alpha\epsilon^2Q(y),\quad P'(y)=(\alpha-1)\epsilon Q(y)+\mathcal{O}(\epsilon^2),\end{align}
so that 
\begin{align}\frac{P'(y)}{P(y)}=-(\alpha^{-1}+1)\frac{1}{\epsilon}+\mathcal{O}(\epsilon^2).\end{align}
The left hand side is finite, as can be checked by explicit computation. This means that $\alpha=-1$, to leading order in $\epsilon$, and 
\begin{align}\tau_{\text{comp.}}=\frac{G}{\ell^{d-1}}\frac{4\pi y^{6-2 d}}{ \left( (d-1) \Omega  \left(d y^2+d-2\right)\right)}\frac{1}{\epsilon}.\end{align}
Again, near extremality the qualitative behavior is similar to the four-dimensional case, since the powers of $\epsilon$ do not change.

We are also interested in the opposite regime, far from extremality. We get
\begin{align}\tau_{\text{orth.}}^2&\approx\frac{G}{\ell^{d-1}} \frac{64 \pi ^2 s^{3-d} \left(d \left(s^2-1\right)+2\right)}{(d-1) \Omega  \left(d s^2+d-2\right)^3},\nonumber\\\tau_{\text{comp.}}&=\tau_{\text{ML}}=\frac{\pi}{2M}=\frac{G}{\ell^{d-1}}\frac{8 \pi ^2 }{(d-1) \Omega_{d-1}  \left(s^{d-2}+s^d\right)}.\end{align}
Just like in the three-dimensional case, \eq{lloyd} is saturated \cite{Cai:2016xho}, but we have $\tau_{\text{comp.}}\ll \tau_{\text{orth.}}$. 
\subsection{Fixed $\mu$ ensemble}
In the grand canonical ensemble, we have instead
\begin{align}\tau_{\text{orth.}}=\frac{1}{\sqrt{-(\beta G)''}},\end{align}
where $G$ is the Gibbs free energy of the charged black hole, which can be found e.g. in \cite{Chamblin:1999tk}. It is
\begin{align}G=\frac{\Omega_{d-1}}{16\pi G \ell^2}\left(\ell^2r^{d-2}\left((1-\frac{2(d-2)}{d-1}\mu^2\right)-r^d\right)\end{align}
while \eq{betis} in terms of the chemical potential is now
\begin{align}\beta =\frac{4 \pi  r}{d r^2+(d-2) \left(1-\frac{(2 (d-2)) \mu ^2}{d-1}\right)}.\end{align}
With these, we can compute
\begin{align}\tau_{\text{orth.}}^{-2}&=\left(\frac{d\beta}{dr}\right)^{-1}\frac{d}{dr}\left(\frac{dG}{d\beta} \left(\frac{d\beta}{dr}\right)^{-1} \right)\nonumber\\&=-\frac{\Omega_{d-1}  r^{d-3} \left((d-2) \ell^2 \left(2 (d-2) \mu ^2-d+1\right)-(d-1) d r^2\right)^3}{64 \pi ^2 (d-1) G \ell^4 \left((d-2) \ell^2 \left(2 (d-2) \mu ^2-d+1\right)+(d-1) d r^2\right)}.\end{align}
This expression simplifies considerably in terms of the extremal black hole radius $y\ell$ (we have also introduced again $r\equiv s\ell$),
\begin{align}\tau_{\text{orth.}}^{2}&=\frac{G}{\Omega_{d-1}\ell^{d-3}}\frac{64 \pi ^2 \left(s^2+y^2\right)}{(d-1) d^2   \left(s^2-y^2\right)^3 s^{d-3}}.\end{align}
Similarly, in the grand canonical ensemble, the ML time is now computed using $(M-\mu Q)- (M-\mu Q)_{0}$, rather than just $M-M_0$. We get 
\begin{align}\tau_{\text{ML}}=\frac{G}{\ell^{d-2}\Omega_{d-1}} \frac{8 \pi ^2 (d-2)s^2}{(d-1)   \left(s^d \left((d-2) s^2-d y^2\right)+2 s^2 y^d\right)}.\end{align}
We are interested in the near-extremality expressions:
\begin{align}\tau_{\text{orth.}}&=\sqrt{\frac{G}{\Omega_{d-1}\ell^{d-3}}}\frac{1}{\sqrt{d^2(d-1) y^{d-2}}}\frac{1}{\epsilon^{3/2}}+\mathcal{O}(\epsilon^{-1/2}),\nonumber\\ \tau_{\text{ML}}&=\frac{G\Omega_{d-1}}{\ell^{d-2}\Omega_{d-1}}\frac{8 \pi ^2}{(d-1) d y^{d-2} }\frac{1}{ \epsilon ^2}+\mathcal{O}(\epsilon^{-1}).\end{align}
We see that the $\epsilon$ scalings near extremality are the same as in the canonical ensemble, just as we advertised in the main text.

\end{appendices}

\providecommand{\href}[2]{#2}\begingroup\raggedright\endgroup

\end{document}